\documentclass{an}
\usepackage{graphicx}
\usepackage{times}
\usepackage{fancyhdr}
\begin{document}

\title{Searching for proto-brown dwarfs: Extending near IR spectroscopy of protostars below the hydrogen burning limit\footnote{The data presented herein were obtained at the W.M. Keck Observatory, which is operated as a scientific partnership among the California Institute of Technology, the University of California and the National Aeronautics and Space Administration. The Observatory was made possible by the generous financial support of the W.M. Keck Foundation.}}

\author{Kevin~R.~Covey$^{1}$, Thomas~P.~Greene$^{2}$, Greg~W.~Doppmann$^{3}$, Charles~J.~Lada$^{4}$, Bruce~A.~Wilking$^{5}$}
\institute{1: University of Washington, Department of Astronomy, Box 351580, Seattle, WA 98195 \\ 
2: NASA Ames Research Center, Mail Stop 245-6, Moffett Field, CA 94035-1000 \\
3: Gemini Observatory, Southern Operations Center, Association of Universities for Research in Astronomy, Inc., Casilla 603, La Serena, Chile \\
4: Harvard-Smithsonian Center for Astrophysics, 60 Garden Street, Cambridge, MA 02138 \\
5: University of Missouri-St. Louis, Department of Physics and Astronomy, 1 University Boulevard, St. Louis, MO 63121}
\date{Received; accepted; published online}

\abstract{Recent observations of nearby star forming regions have offered evidence that young brown dwarfs undergo a period of mass accretion analogous to the T Tauri phase observed in young stars.  Brown dwarf analogs to stellar protostars, however, have yet to be definitively observed.  These young, accreting objects would shed light on the nature of the dominant brown dwarf formation process, as well as provide ideal laboratories to investigate the dependence of the accretion mechanism on protostellar mass.  Recent near infrared surveys have identified candidate proto-brown dwarfs and characterized low mass protostars in nearby star forming regions.  These techniques allow near infrared spectra to diagnose the effective temperature, accretion luminosity, magnetic field strength and rotation velocity of young low mass stars across the stellar/substellar boundary.  The lowest mass proto-brown dwarfs (M$ < $40 M$_{Jup}$), however, will prove challenging to observe given current near IR observational capabilities.
}

\correspondence{covey@astro.washington.edu}

\maketitle

\section{Introduction}

The past decade has witnessed a rapid growth in the number of known sub-stellar objects, or brown dwarfs.  Intense study of these brown dwarfs have extended our understanding of galactic stellar populations and the star formation process beyond the stellar/substellar boundary.  It now appears that the initial mass function (IMF) extends continuously to the substellar regime (Najita et al. 2000), though changes in the exact form of the IMF may be evident at the lowest masses (Muench et al. 2002) or as a function of environment (Luhman 2004).  Young brown dwarfs also appear to pass through an evolutionary stage analogous to the Classical T Tauri phase, harboring circumstellar disks (Jayawardhana et al. 2003, Liu et al.  2003, Luhman et al. 2005), exhibiting emission lines indicative of active mass accretion (Muzerolle et al. 2003, Natta et al. 2004, Mohanty et al. 2005) and displaying photometric variability indicative of stochastic mass accretion (Scholz et al. 2005).  Multiple physical mechanisms have been proposed as the dominant source of sub-stellar objects, such as formation in planetary disks (Papaloizou et al. 2003), ejection from unstable multiple systems (Reipurth \& Clarke 2001) and/or formation from molecular cores with substellar masses (Padoan et al. 2002).   Each model makes differing predictions regarding the presence of sub-stellar disks and mass accretion at ages comparable to those of the Classical T Tauri Stars (CTTS; t $\sim$ 1 Myr).  

Despite these efforts, however, a complete description of the formation of these lowest mass objects remains elusive.  While studies of T Tauri stars and their brown dwarf analogs can reveal much about the products of the star formation process, they provide only tantalizing clues to the physics which dominate the main accretion phase of stars and brown dwarfs, long since completed by the T Tauri stage.  Protostars are commonly thought to be highly reddened, heavily embedded stars undergoing a high level of active mass accretion; observationally, these objects are identified as sources with broad SEDs peaking between 10-100 $\mu$m embedded in dense molecular cores (Lada 1987).  Observations of these protostars, or proto-brown dwarfs in the very low mass regime, will be key to describing the physics which govern the star and brown dwarf formation processes.   

Surveys of embedded protostars have been carried out at optical wavelengths (Kenyon et al. 1998, White \& Hillenbrand 2004).  Indeed, White \& Hillenbrand measured spectral types of M5.5 or M6 for 3 Class I objects in Taurus (04158+2805, 04489+3042 and 04248+2612) on the basis of optical spectroscopy; adopting a dwarf temperature scale for these objects implies masses near or below the substellar limit.  Infrared spectra of two of these objects were obtained in a survey of nearby star forming regions by Doppmann et al. (2005; described in detail in section 2).  Visual inspection of these spectra confirm an M6 spectral type for 04158+2805, while 04489+3042 appears to have an earlier M4 spectral type.  

The extremely reddened nature of embedded, accreting protostars, unfortunately, make optical observations largely inaccessible for all but the closest star formation regions.  Over the past decade, however, advances in infrared detector technology and the rapid proliferation of 8-meter class telescopes have enabled high resolution spectroscopic studies of protostars and proto-brown dwarfs.  In this contribution we discuss recent findings of near infrared surveys of embedded protostars, and assess the opportunity this technique presents for the identification and characterization of proto-brown dwarfs.


\section{Near Infrared Spectroscopic Surveys of Embedded Protostars}

Over the past three years, several groups have used moderate to high resolution near infrared spectrographs to study embedded Class I or flat spectrum protostars (Greene \& Lada 2002, Ishii et al. 2004, Nisini et al. 2005, Doppmann et al. 2005).  These studies have revealed photospheric absorption features in nearly 50 Class I or flat spectrum protostars.  The bulk of these protostars (42) have been observed by Doppmann et al. (2005; hereafter D05) in a survey of Class I and flat spectrum protostars in the Taurus, Serpens, and $\rho$ Ophiuchi star forming regions.  The results of this survey, containing a homogenous sample with more than 80\% of the protostars with currently measured near-IR absorption features, provide a glimpse of the utility of this technique for studying highly embedded protostars and proto-brown dwarfs.  

An assessment of the evolutionary state of Class I and flat spectrum protostars is a key observational goal of many protostellar studies.  D05 observed selected regions in the K band ( $\sim$2.3 $\mu$m) containing a number of prominent atomic Al, Mg, Na, Ti, \& Fe lines, along with a strong molecular CO bandhead, that are seen in absorption for late type stars.  Comparison of observed near IR spectra with a grid of synthetic model spectra (Sneden 1973, Hauschildt et al. 1999) allow the determination of protostellar photospheric effective temperatures (T$_{eff}$s).  Spectroscopic analysis of photospheric lines  reveal the presence of excess continuum flux; this emission is thought to arise from circumstellar material heated by accretion, typically known as veiling.  The amount of veiling flux is typically expressed as a ratio of excess continuum flux to photospheric flux, r$_K$; typical protostars possess r$_K \sim$1.8.  Veiling measurements, combined with near IR photometry, provide estimates of absolute photospheric and accretion luminosities, allowing these sources to be placed on an HR diagram (see Figure 1).  While comparison between the location of observed sources and model tracks in an HR diagram are useful for understanding the age and mass of a protostellar population, significant uncertainties in extinction corrections for these heavily embedded sources currently prevent a clear separation of Class I sources and CTTSs in the HR diagram (compare Fig. 1 to Figure 15 of D05).  Detailed modeling of protostellar envelopes on a source by source basis may be required to derive confident luminosity estimates.

\begin{figure}
\resizebox{\hsize}{!}
{\includegraphics[]{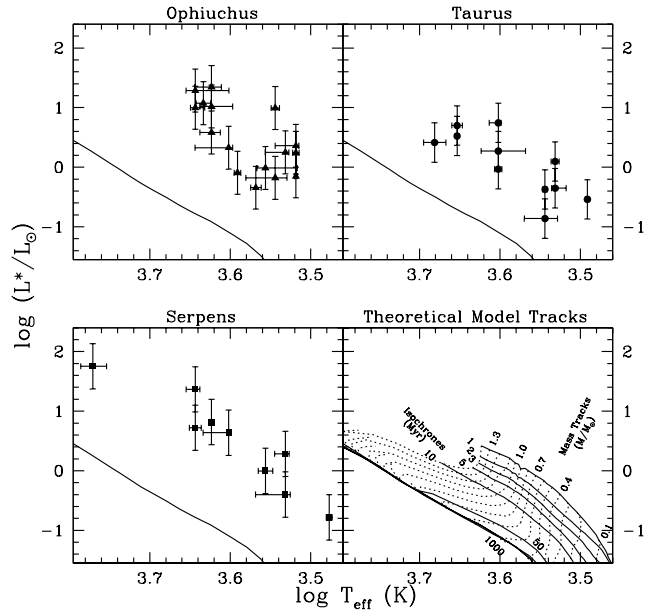}}
\caption{Photospheric effective temperatures and luminosities (explicitly neglecting accretion components) measured for Class I and flat spectrum protostars by D05.  Separate HR diagrams are shown for sources from each of three nearby star formation regions: $\rho$ Ophiuchus, Taurus and Serpens.  Theoretical pre-main sequence tracks calculated by Baraffe et al. (1998) are shown in the bottom right panel for comparison.  The 1 Gyr isochrone from the Baraffe models are shown for reference as a solid line in the $\rho$ Ophiuchus, Taurus and Serpens HR diagrams. Significant vertical error bars result from uncertain extinction corrections for these heavily embedded objects.  Originally presented as Fig. 14 in D05.}
\label{observationalHR}
\end{figure}

Velocity information is encoded in the position and breadth of photospheric absorption lines, which can be analyzed to reveal the radial and rotational velocity of the protostellar photosphere.  Analysis has revealed significant differences in the rotation velocities of Class I and CTTS sources, suggesting that angular momentum is extracted during the protostar's evolution, possibly through magnetic interactions between protostars and their circumstellar disks (Covey et al. 2005).  The radial velocities of protostars and their associated CO gas also constrain the amount of dynamical evolution which could have occurred between the protostellar and CTTS phases.   The D05 sample indicates the protostellar 1-dimensional velocity dispersion must lie below the 3 km s$^{-1}$ level and is fully consistent with the dynamical state of CTTSs, in opposition to what might be expected in the ejection model of brown dwarf formation (Covey et al. 2006).  

The physical state of circumstellar material can also be investigated through the analysis of near IR spectra.   More than half the protostars (34/52) observed by D05 display HI Brackett $\gamma$ in emission, indicative of significant levels of mass accretion, while slighly less than half of the sample (23/52) show H$_{2}$ emission, indicative of outflow shocks.   The exact shape of these emission line features also contain information about the exact velocity structure of the accretion and outflow phenomena.  Approximately 15\% of the protostars observed by D05 also show CO and Na features in emission, indicative of emission from the irradiated surface of a circumstellar disk.  

Even sources without detected photospheric absorption features provide insight into the nature of their protostellar cores, as either rapid rotational velocities or high levels of accretion luminosity are necessary to weaken lines below detectability in high quality (S/N $\sim$ 100) spectra (see Figure 2).  A number of low luminosity sources in the D05 survey which failed to reveal photospheric absorption features represent possible proto-brown dwarfs.  EC103, for example, has a flat continuum spectrum that implies a veiling luminosity which is a factor of three or greater than the photospheric luminosity in the K band.   The bolometric luminosity of EC103, however, is only 0.24 L$_{\odot}$ -- attributing a significant fraction of this luminosity to accretion implies a very low mass protostellar core.  

\begin{figure}
\resizebox{\hsize}{!}
{\includegraphics[]{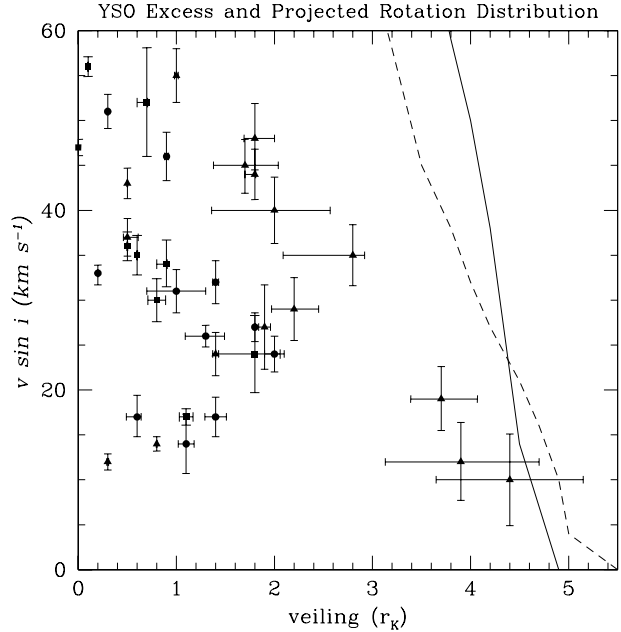}}
\caption{Rotation velocities ($v$ sin $i$s) and veiling (r$_K$) for the 42 protostars with detected photospheric absorption features in the D05 survey.  Shown as a solid and dashed line are estimates of the sensitivity limit of the D05 observations.  Objects above and to the right of these lines have features too weak to be detected by S/N $\sim$ 100 spectra.  Originally presented as Fig. 12 in D05.}
\label{detectionlimits}
\end{figure}

\section{Prospects for Spectroscopic Detection of Proto-Brown Dwarfs}

Observations of protostars with sub-stellar masses would provide ideal laboratories for understanding the formation of very low mass objects.  The very existence of isolated sub-stellar protostars provides a key observational test for formation mechanisms involving dynamical interactions with companion stars.  While the featureless protostars described above may represent such objects, the most valuable laboratory for studying proto-brown dwarf formation will require the detection of protostellar absorption features indicative of a very cool photosphere.   

Early results from Spitzer Space Telescope IRAC imaging of the Elephant Trunk Nebula (750 pc; Reach et al. 2004) provide an example of how Spitzer data can be used to identify low luminosity proto-brown dwarf candidates.  Three Class I objects are detected with L$_{MIR} < 0.2 L_{\odot}$, with the least luminous source (L$_{MIR} = 0.1 L_{\odot}$) undetected at K band in 2MASS; a typical K - [3.6] color for other Class I sources in the association predicts K $\sim$ 17.  Spitzer surveys of nearby star formation regions (e.g., $\rho$ Ophiuchus, 140 pc) should easily detect sources of similar luminosity to that described above, as well as numerous intrinsically fainter sources that span the stellar/sub-stellar boundary.  Due to their relative proximity, these objects will have K mag $\sim$ 14.5.  Models of broadband SEDs, however, possess significant degeneracies between structural parameters (disk inclination, grain properties, accretion luminosity, brown dwarf age \& mass, and non-isotropic extinction/scattering in proto-brown dwarf envelopes); only detailed spectroscopic confirmation and study can conclusively identify and probe the physics of these forming brown dwarfs (Luhman 2005).  Perhaps more fundamentally, near IR spectroscopic follow-up may also be necessary to confirm that these sources are truly embedded protostars and not reddened background stars or IR luminous galaxies.  

These sources present attractive candidates for moderate resolution, near infrared spectroscopic follow up.  Current spectrographs on 8 meter class telescopes (NIRSPEC, IRCS, NIRI, GNIRS, ISAAC), for example, will achieve S/N ratios of $\sim$ 50 in one hour of integration time at R $\sim$ 1000-4000 for a source with K$\sim$14-15.  These spectra will provide valuable constraints on the physical properties of young brown dwarfs, similar to those provided by our recent work on higher mass protostars.  Several temperature sensitive near-IR spectral features have been identified in both the K (Na I, Mg I, Ca I, CO; Ali et al. 1995, Wallace \& Hinkle 1997) and L (SiO, OH, Mg I; Wallace \& Hinkle 2002) bands.  Past calibrations indicate these features can be used at moderate resolution to predict photospheric temperatures to accuracies of $\pm$ 300 K (see Figures 3 \& 4).  Estimates of surface gravity would produce another measure of the current physical state of these objects.  Previous low resolution work produced estimates of log $g$ to $\pm$ 0.5 dex through analysis of alkali metal lines in the K band (Gorlova et al. 2003), while spectra of giant stars at moderate resolution show a clear luminosity signature using an index formed from K band CO, Na, and Ca features (Ramirez et al. 1997).  Additional information will be available from analysis of particular line strengths, particularly mass accretion rates from Br $\gamma$ and veiling luminosity from photospheric absorption line depths.

\begin{figure}
\resizebox{\hsize}{!}
{\includegraphics[]{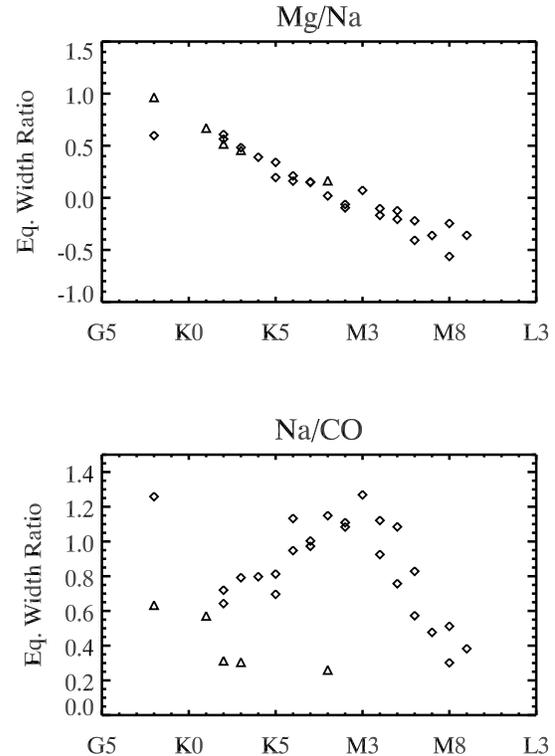}}
\caption{Temperature sensitive absorption line ratios as a function of spectral type, as measured from the spectral standards observed by D05, degraded to moderate resolution (R $\sim$ 2000).  Ratios of absorption features are particularly useful for objects with ongoing accretion, as they are insensitive to the presence of accretion luminosity.  \textit{Top:} A ratio of absorption in a wavelength window centered on the 2.1066 $\mu$m (containing a Mg I line at warmer temperatures and complex continuum
structure at cooler temperatures) to Na I absorption at $\lambda \sim
2.208 \mu$m.  Negative values of this ratio for later
spectral types are due to negative equivalent widths measured for Mg I.  
These negative values do not indicate, however, the detection of an
actual Mg I emission line, but rather the growth of a strong absorption
figure (Line X; see Fig. 4) in the wavelength region used to set the      
continuum level for this measurement.  As Line X increases in strength
towards cooler temperatures, the continuum level in this area becomes 
increasingly underestimated, resulting in the non-physical measurement of
Mg I `emission' which nonetheless serves as a reliable temperature
indicator.  \textit{Bottom:} A ratio of the Na doublet absorption to the
CO absorption bandhead strength at 2.3 $\mu$m.  Ratios for main sequence  
stars are shown as diamonds, while ratios for giant (lum. class III) stars
are shown as triangles.  Note that both ratios are sensitive to
temperature changes at the canonical pre-main sequence stellar/sub-stellar
boundary ($\approx$ M6).}
\label{detectionlimits}
\end{figure}

\begin{figure}
\resizebox{\hsize}{!}
{\includegraphics[]{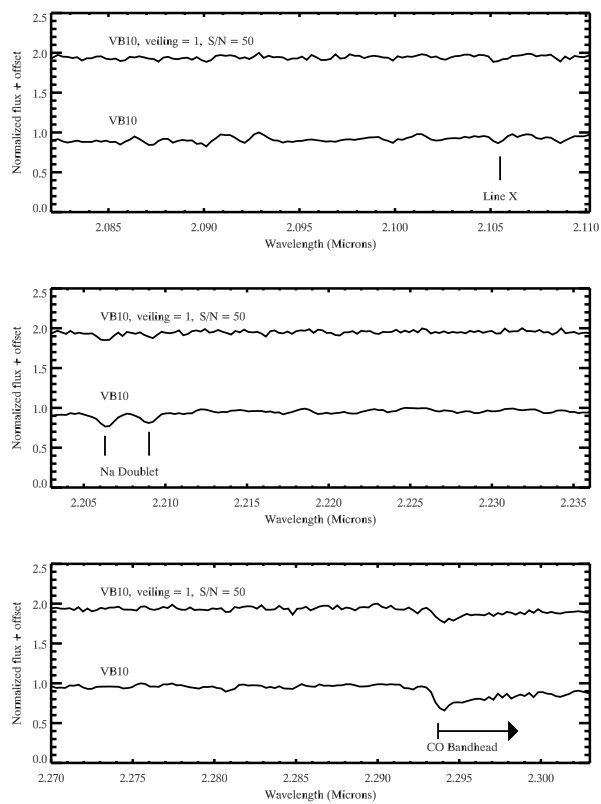}}
\caption{An example of the features which will be present in spectra of proto-brown dwarfs.  Shown in each panel is a segment of the K band spectrum of the M8 dwarf VB10, originally observed by D05 and subsequently degraded to moderate resolution (R $\sim$ 2000).  The lower spectrum of each panel shows an unaltered spectrum of VB10 (save for the degraded resolution), while the higher spectrum simulates a spectrum which would be observed from a proto-brown dwarf.  This spectrum is constructed from observations of VB10 with the addition of moderate veiling flux (r$_K$ = 1) and random gaussian noise (S/N = 50).  Prominent spectral features are indicated in each panel, including a strong feature at 2.105 $\mu$m seen in late type stars that we have identified as 'Line X'.  The presence of this line, as well as numerous absorption features across the continuum at these cool temperatures, leads to the negative Al/Mg absorption line strengths measured for the coolest stars in Figure 3.  Actual moderate resolution observations of proto-brown dwarfs will cover significantly larger regions of the spectrum than are shown here, including a prominent temperature and gravity sensitive Ca I line.}
\label{observationalHR}
\end{figure}

Characterizing a significant sample of young brown dwarfs will provide insight into a number of interesting scientific problems.  Coupled estimates of T$_{eff}$ and surface gravity allow observations to test theoretical isochrones under the assumption of coeval populations.  Binary systems discovered during AO integral field efforts may provide dynamical mass estimates for detailed comparison with predictions of pre-main sequence models.  Mass accretion rate measurements will allow extension of the host mass/mass accretion rate relation into the substellar regime at extreme youth, as well as provide a laboratory to probe the time dependence of accretion through comparison to CTTS results.  Assuming protostars follow largely vertical Hayashi tracks as they collapse to the main sequence, stellar masses can be estimated based solely on observed T$_{eff}$s, and the first proto-brown dwarf mass function can be constructed.  Bright targets with particularly strong features will be ideal candidates for high resolution followup, providing accurate rotation measurements to probe angular momentum evolution, as well as identifying slow rotators whose magnetic field strengths can be derived from Zeeman splitting of absorption lines (Johns-Krull et al. 1999).  Lastly, sources with clear disk emission signatures can be observed at high spectral and spatial resolution to map out parameter space for disk properties (kinematic and temperature profile, inclination, etc.).

Determining physical parameters of objects well below the sub-stellar boundary, however, may prove difficult given current observational capabilities.  The pre-main sequence models of Baraffe et al. (1998) predict a 40 M$_{Jup}$ brown dwarf in Taurus will have a K magnitude $\sim$ 12.4 at 1 Myr.  Extinctions typical of Class I sources in Taurus are A$_K \sim$ 3 magnitudes, placing a potential proto-brown dwarf at m$_K \sim$ 15.5.  Observing objects at this magnitude and fainter with current near infrared spectrographs will require very long integration times to achieve modest signal-to-noise ratios.  Possible observational techniques that may allow detailed study of extremely low mass proto-brown dwarfs are:

\begin{itemize}
\item{Multi-object capability -- observatories with strongly multiplex near infrared spectrographs (e.g., FLAMINGOS) allow the assembly of reasonable samples of objects requiring extremely long integration times, as numerous objects can be observed simultaneously.}
\item{Lower resolution studies -- Strong, broad water absorption features useful for assessing effective temperature are known to exist in near IR spectral windows.  Lower resolution observations will be less productive scientifically than moderate/high resolution observations ($v$ sin $i$, $v_{rad}$ and log $g$ measurements will be difficult or impossible, T$_{eff}$ and r$_K$ become degenerate at low resolution), but may serve to identify interesting objects for dedicated followup studies.}
\item{L band observations -- Cool, heavily reddened proto-brown dwarfs should release more flux in the L band than in shorter wavelength windows.  If spectral features sensitive to temperatures appropriate for very cool proto-brown dwarfs can be identified, observations in the L band may return higher S/N ratios for a given exposure time than at K band.}
\end{itemize}

\section{Summary}

Near infrared surveys have begun to provide detailed diagnostics of the physical state (T$_{eff}$, log $g$, r$_{K}$, v$_{rot}$, v$_{rad}$) of moderate samples of heavily embedded protostars.  These techniques are providing insights into the physics governing mass accretion and angular momentum transfer during the protostellar phase; coupling current techniques with detailed extinction modeling or sensitive surface gravity indicators will allow more precise estimate of a protostar's evolutionary state.  Though these techniques have not yet identified any confirmed substellar protostars, current observational limits should allow detection of proto-brown dwarfs by observations at moderate resolution (R $\approx$ 2000).  Delving deep below the sub-stellar limit, however, may prove challenging given current observational limitations.

\acknowledgements

The authors wish to recognize and acknowledge the very significant cultural role and reverence that the summit of Mauna Kea has always had within the indigenous Hawaiian community.  We are most fortunate to have the opportunity to conduct observations from this mountain.  All data have been reduced using IRAF; IRAF is distributed by the National Optical Astronomy Observatories, which are operated by the Association of Universities for Research in Astronomy, Inc., under cooperative agreement with the National Science Foundation.  This research has made use of NASA's Astrophysics Data System Bibliographic Services, the SIMBAD database, operated at CDS, Strasbourg, France, and the VizieR database of astronomical catalogues (Ochsenbein et al. 2000).  K.R.C gratefully acknowledges the support of NASA grant 80-0273. All authors acknowledge support from NASA Origins of Solar Systems program via UPN 344-39-00-09 while conducting this work.


\end{document}